\documentclass[fleqn,usenatbib]{mnras}

\usepackage[T1]{fontenc}

\DeclareRobustCommand{\VAN}[3]{#2}
\let\VANthebibliography\thebibliography
\def\thebibliography{\DeclareRobustCommand{\VAN}[3]{##3}\VANthebibliography}

\usepackage{graphicx}	
\usepackage{amsmath}	
\usepackage{amssymb}	
\usepackage{longtable}

\title[Dichotomy between BH and NS LMXBs]{A Temporal Scale to Track the Spectral Transitions in Low-Mass X-ray Binaries}
\author[E. Sonbas et al.]{
E. Sonbas,$^{1,3}$\thanks{E-mail: edasonbas@gmail.com}
K. Mohamed,$^{2,3}$
K. S. Dhuga,$^{3}$
and E. G\"o\u{g}\"u\c{s}$^{4}$
\\
$^{1}$Department of Physics, Adiyaman University,  02040 Adiyaman, Turkey\\
$^{2}$Department of Physics, Faculty of Science, Sohag University, Sohag, 82524, Egypt. \\
$^{3}$Department of Physics, The George Washington University, Washington, DC 20052, USA\\
$^{4}$Faculty of Engineering and Natural Sciences, Sabanc\i~University, Orhanl\i~- Tuzla, Istanbul 34956, Turkey\\
}

\date{Accepted 2021 November 16. Received 2021 November 16; in original form 2021 January 27}

\pubyear{2020}

\begin{document}
\label{firstpage}
\pagerange{\pageref{firstpage}--\pageref{lastpage}}
\maketitle

\begin{abstract}
\noindent
The results of a temporal analysis of observations for a sample of nine low-mass X-ray binaries (LMXBs) are presented. Of these sources, five host a neutron star (NS) primary (4U1608-52, Aql X-1, 4U1705-44, GX17+2 and Cyg X-2), and four host a black hole (BH) (GX339-4, XTE J1859+226, H1743-322 and MAXI J1659-152). The NS group includes three Atolls and two Z-type sources. We utilized archival Proportional Counter Array (PCA)/RXTE data to construct high-resolution lightcurves. A wavelet transform of the lightcurves is deployed to extract a minimal time scale (MTS) associated with the spectral state of the sources. The MTS, together with the fractional root-mean-square (RMS) and hardness ratios, is used to construct RMS-MTS and hardness-MTS diagrams that enable a direct comparison of the evolution of spectral transitions in the target sources. Observations with high fractional RMS and high hardness cluster in a broad region occupied jointly by BH and NS sources. For low fractional RMS observations the Atolls exhibit large MTS whereas Z-type sources exhibit small MTS. This new feature raises the possibility of discriminating between these two types of sources. Moreover, in the hardness-MTS plane, BH sources are the sole occupiers of the low-hardness and small-MTS domain thus potentially signaling a unique property for distinguishing BH and NS hosts in LMXBs.  
\end{abstract}

\section{Introduction} \label{sec:intro}
\noindent
It is now widely accepted that black hole (BH) low-mass X-ray binaries (LMXBs) recurrently follow a similar evolution when they undergo outbursts, displaying various spectral states likely connected to different accretion regimes onto the BH (e.g. \citet{1992ApJ...391L..21M}, \citet{1993ApJ...403L..39M}, \citet{2006csxs.book...39V}, \cite{2004MNRAS.355.1105M}, \citet{2011BASI...39..409B},  \citet {2018MNRAS.481.3761G}, \citet{2021MNRAS.502.1334S} and references therein). Of these spectral states, there is general consensus regarding the two most prominent ones: the Low/Hard (LHS) state and the High/Soft state (HSS). In addition, intermediate states \citep{2007A&ARv..15....1D, 2005Ap&SS.300..107H,  2010MNRAS.403...61D} are also often observed, and along with the main states, typically trace out an anticlockwise-evolving  q-shape trajectory in the hardness-intensity diagram (HID; \citet{2001ApJS..132..377H}, \citet{2005A&A...440..207B}, \citet{2006ARA&A..44}, and also see \citet{2012A&A...542A..56N}, \citet{2018MNRAS.480.4443T} and references therein).\\ 
\\
LMXBs, hosting a neutron star (NS), are also known to undergo occasional outbursts that can last anywhere from weeks to many months, before eventually returning to a low-luminosity quiescence state. During the outbursts  \citep{2003MNRAS.338..189M, 2009ApJ...696.1257L, 2010ApJ...719..201H}, these sources too exhibit spectral patterns and transitions reminiscent of several states; a hard state, thought to be dominated by Compton scattering off hot electrons; a soft state, associated with thermal emission from the accretion disk, and an intermediate state that exhibits both thermal and non-thermal emission features. The NS-LMXBs as a distinct group, has in the past, been divided into two sub-groups according to the shape and tracks in the HID and the evolution of the softness/hardness of the source on the color-color diagram (CCD; \citet{1989A&A...225...79H}). At high accretion rate $(L_X > 0.5 L_{Edd})$ some sources (classified as Z-type sources) display Z-shaped tracks, while lower accretion-rate systems $(0.01 L_{Edd} < L_X < 0.5 L_{Edd})$ show a different phenomenology and are classified as Atoll sources based on the tracks displayed in the CCD. Some later studies \citep{2002MNRAS.331.47, 2002ApJ...568L..35M}, on the other hand, have strongly suggested that the two sub-groups are in fact very similar in their evolution as a function of the accretion rate questioning the apparent distinction based on the CCD tracks. Nonetheless, studies of the spectral tracks, both in the HID and the CCD, show hysteresis-like loop structures for most of the LMXBs whether they host a BH or a NS (e.g. \citet{2010MNRAS.403...61D}, \citet{2001ApJS..132..377H}, and \citet{2006MNRAS.367.1113B}). The transition tracks do not necessarily follow well-defined contours, instead they exhibit considerable dispersion, forming band-like patterns (loop-like structures) on the HIDs as the intensity of the sources varies from outburst to outburst. Indeed, variation is observed within individual sources. \\
\\
In addition to hardness ratios, the fractional root-mean-square (RMS), has proven to be a useful tool in tracking the spectral states and their transitions. The RMS is determined by integrating the power spectral density (PSD) in a specific frequency range, see (\citet{1983ApJ...266..160L}, \citet{1988SSRv...46..273L}, \citet{1989ARA&A..27..517V}, \citet{1990A&A...230..103B}, and \citet{1992ApJ...391L..21M}), and the hardness is usually taken to be the ratio of counts in two adjacent spectral bands nominally depicting the hard and soft bands respectively. In a study of a sample of NS-LMXBs, \cite{2014MNRAS.443.3270M} deployed HIDs and the RMS intensity diagram (RID) to demonstrate that hysteresis loops, qualitatively similar to those observed in BH binaries, is a common feature in the persistent and the transient NS-LMXB population when accreting at moderate rates. At higher accretion rates, these patterns disappear and the NS binary systems remain in a thermal dominated state characterized by flaring behavior and fast color changes. Their study also indicated that systems located in the same region of the RID may share similar phenomenology, thus suggesting a potential common framework in the interpretation of BH and NS accretion states.\\
\\
Recently, \cite{2020MNRAS.853.150S} introduced the `minimal time scale' (MTS), a temporal property, to track the spectral transitions via the intensity variability diagram (IVD) which maps the count rates of a source as a function of the MTS as the source undergoes spectral evolution during an outburst. The MTS represents the time scale associated with the shortest-duration temporal feature in a lightcurve or equivalently the highest frequency component of the signal above Poissonian noise in a PSD \citep{2021arXiv210104201M}. A robust method for extracting the MTS is presented by \citet{2013MNRAS.432..857M, 2013MNRAS.436.2907M}, in which the authors used a technique based on wavelets to analyze a sample of gamma-ray bursts and demonstrated the separation of white noise (background) from red noise (signal) at a level of a few milliseconds. In their study of the BH transient GX339-4, \cite{2020MNRAS.853.150S}, clearly demonstrated that the IVD is also sensitive to the hysteresis loops and the q-shape of the evolutionary tracks observed in the more familiar HIDs (and the RIDs). In addition, they also reported a positive correlation between the fractional RMS and the MTS albeit with considerable scatter. In a follow-up study of two NS  binaries, namely 4U 1605-68 and Aql X1, the same team \citep{2021arXiv210104201M} reported an anti-correlation between RMS and MTS. It was speculated that the difference could be due to the fact that NS binaries do not exhibit very low hardness ratios that are observed in BH transients such as GX339-4 in the high/soft state. The presence of the hard surface in the case of neutron stars was mentioned as the prime reason \citep{2003MNRAS.342..361N}. Nonetheless, taken together, the HID, the RID, and now the newly devised IVD, featuring the temporal parameter MTS, offer an array of diagnostic tools in probing the underlying accretion process, and the configurational coupling of the accretion disk to the corona surrounding the compact object. Of course, understanding the roles of the accretion disk and the corona, and disentangling their respective contributions to the observed emission, remains one of the primary aims in the continued study of spectral and temporal properties of LMXBs.\\
\\
Our motivation in this study is to further explore and quantify the utility of the IVD in tracking the spectral changes that occur in a number of well studied BH and NS (Atoll and Z-type) binaries. In addition, we wish to map the response of these sources in the RMS-MTS plane, and determine whether it reveals the existence of a dichotomy between these classes of sources. The layout of the paper is as follows: in section 2, we describe the data selection, the reduction procedures, the extraction of the needed high-resolution lightcurves, the construction of the PSDs and the determination of the fractional RMS. The basic features of the wavelet technique used to extract the MTS are also noted in this section. In section 3, we present our main results, the behavior of the IVDs, the nature of the RMS-MTS correlations for the sample sources, and the discussion of the relative outcomes for our sample sources. We present the summary and the conclusions of our study in section 4.
\section{Data Reduction and Methodology}
\noindent
In this study, we analyzed publicly available RXTE/PCA observations of 9 LMXBs: 5 that host a NS primary (4U1608-52, Aql X-1, 4U1705-44, GX17+2 and Cyg X-2), and 4 that host a BH (GX339-4, XTE J1859+226, H1743-322 and MAXI J1659-152). The NS group includes 3 Atolls (Table 1) and 2 Z-type sources (Table 2). In the case of the Atoll source 4U1608-52, the specific observations covered the 2002 and 2007 outbursts; for Aql X-1 (another Atoll), the data covered the 1999 and 2000 outbursts. For the third Atoll (4U1705-44), we analyzed the 2007 outburst. For the Z-type sources (GX17+2 and Cyg X-2), we analyzed the 1999 outbursts. For the BH sources (see Table 3), the observations covered the 2002 and 2010 outbursts for GX339-4 (see \cite{2020MNRAS.853.150S}), the 1999 outburst for XTE J1859+226,  2010 outburst for MAXI J1659-152, and the 2003 outburst for H1743-322. These sources are good test candidates because they are well studied and the majority are known to exhibit a range of spectral transitions (\citet{2018MNRAS.477.5220P}, \citet{2018MNRAS.477.5437A}, \citet{2018MNRAS.481.3761G}, \citet{2020Ap&SS.365...41A}, \citet{2021MNRAS.502.1856M}, \citet{2021MNRAS.502.1334S}, and references therein).\\
\\
We deployed the standard RXTE \footnote{\tiny $https://heasarc.gsfc.nasa.gov/docs/xte/xhp\_proc\_analysis.html$} data analysis tools of HEASOFT V.6.26. For our analysis we used PCA data modes from different channels with different time resolution including high resolution Good-Xenon or Event data modes that cover the full energy band, otherwise we combined Single-Bit and Event data modes to cover the full energy band. In the cases where we did not have any Good Xenon and Event data to cover the full energy band, there we combined two different modes to cover the full energy band. The primary reason for using the full energy band was to obtain a measure of the background noise which would appear in the highest available channels. For the final analysis, we created lightcurves with a time resolution of $2^{-12}$s (i.e., $\sim244$ $\mu$s). These lightcurves were created using standard screening criteria by applying source elevation of greater than 10 degrees, the pointing offset less than 0.02, and all PCUON. For each observation, we generated a background model for bright objects using the FTOOL (pcabackest). This background model is used in determining the PSD normalization.\\
\\
We followed the procedures described by \citet{2000MNRAS.318..361N}, \citet{2003A&A...407.1039P} and \citet{2005A&A...440..207B}) to construct PSDs for lightcurve segments of 16s duration (and some for durations of 64s for consistency checks) and for Nyquist frequency of 2048 Hz using Powspec 1.0 (Xronos5.22). Fast Fourier transforms were performed for each segment to extract a normalized PSD \citep{1983ApJ...266..160L} for each observation. The segments were averaged and rebinned geometrically (a factor of -1.05) to smooth the power spectrum. We normalized the PSDs following \cite{1989A&A...225...79H, 1990A&A...230..103B,1992ApJ...391L..21M}, taking into account the background renormalization and Poisson level. In principle PSDs can be determined by using lightcurves with or with background subtraction; in the case of background subtracted lightcurves the time binning for those lightcurves is typically the order of 16 seconds. If finer time binned lightcurves are deployed no background subtraction is done for those lightcurves before performing the Fourier transform to generate the PSDs. However, the normalization of each PSD can be adjusted based on the estimate of the background level obtained from the high-energy channels, i.e., this renormalization procedure in a sense `compensates' for the lack of background subtraction at the lightcurve stage. We also tested for deadtime corrections to the PSDs using the procedure outlined by \cite{1999ApJ...510..874N, 1995ApJ...449..930Z, 1996ApJ...469L..29Z, 2000ApJ...530..875J} but found the corrections to be minor (the order of few percent) in the   frequency range of interest. The total RMS was computed directly from the integration of the averaged PSDs in the frequency band (0.1-64 Hz). The uncertainties in RMS were calculated using standard error propagation methods \citep{1989ARA&A..27..517V, 2004A&A...414.1091G}. The hardness ratio was calculated using the channel ranges: 16-35 (6-15 keV) and 0-15 (2-6 keV) respectively. The resulting HIDs provide a visual representation of the loop-like patterns of flux vs. hardness as the sources undergo spectral transitions. The HIDs were used to make sure that we were able to reproduce the results of \citet{2014MNRAS.443.3270M}, and secondly, to identify `complete' loops i.e., isolate those set of observations that indicate a range of spectral transitions for further analysis.\\
\\
For the extraction of the MTS, we deployed a fast wavelet transformation to represent the light curves. Use of wavelets in the analysis of time-series has increased over the years \citep{Flandrin89, Flandrin92, Mallat89, 1993ApJ...411L..91S, 1997ApJ...487..396S, 2005MNRAS.361..645L, 2010A&A...515..65L}. For our analysis, we adopted the methodology originally described by \citet{2013MNRAS.432..857M, 2013MNRAS.436.2907M}. In addition, we refer the reader to the more recent studies by \cite{2020MNRAS.853.150S} and \cite{2021arXiv210104201M} where the general background of the wavelet technique and the essential details are further discussed. In essence, the procedure requires the representation of the lightcurve by a set of basis functions. In our case, we opted for the Haar wavelet \citep{1992ComPh...6..697D}, one of the simplest wavelets available: it is constant over its time interval (i.e., a 'box'-like profile) and similar to the model assumed in the Bayesian block method \citep{2013ApJ...764..167S}. The extracted detail coefficients ($d_{jk}$; see the appendix in \cite{2021arXiv210104201M}) are used to determine the variance, which in turn is used to construct a logscale diagram, a plot of log of variance vs. frequency (octaves).\\
\\
Similar to PSDs in Fourier space, the logscale diagrams (see \citet{2021arXiv210104201M} for an example of such a diagram) provide a measure of the signal plus background; the signal of interest (sometimes referred to as red-noise), appears as a sloped region, whereas the background, a combination of source-related and non-intrinsic noise, commonly referred to as white-noise, appears as a flat region. The intersection of the two regions occurs at some characteristic frequency corresponding to a characteristic timescale that we take to be the minimal timescale. The two respective regions of the logscale are fitted, typically with simple linear functions, to determine the intersection point. The value of the extracted frequency (given in octaves, \textit{j}) is then converted to a timescale by using the binning time of the light curve data (i.e., the MTS (s) $\sim$ 2$^j$ x timebin). As many different wavelets, with a wide variety of time profiles are available, it was deemed necessary to test the extraction procedure and results by using an alternate wavelet function. For the comparison, we used the Mayer wavelet, a function that possesses a significantly different time profile than the Haar, and is readily accessible in the wavelet Toolbox of MATLAB (R2019a). We obtained excellent agreement over the relevant time scales.

\section{Results and Discussion}
\begin{figure}
	\hspace{-1.1cm}
	\centering      
	\includegraphics[scale=0.33, angle = 0]{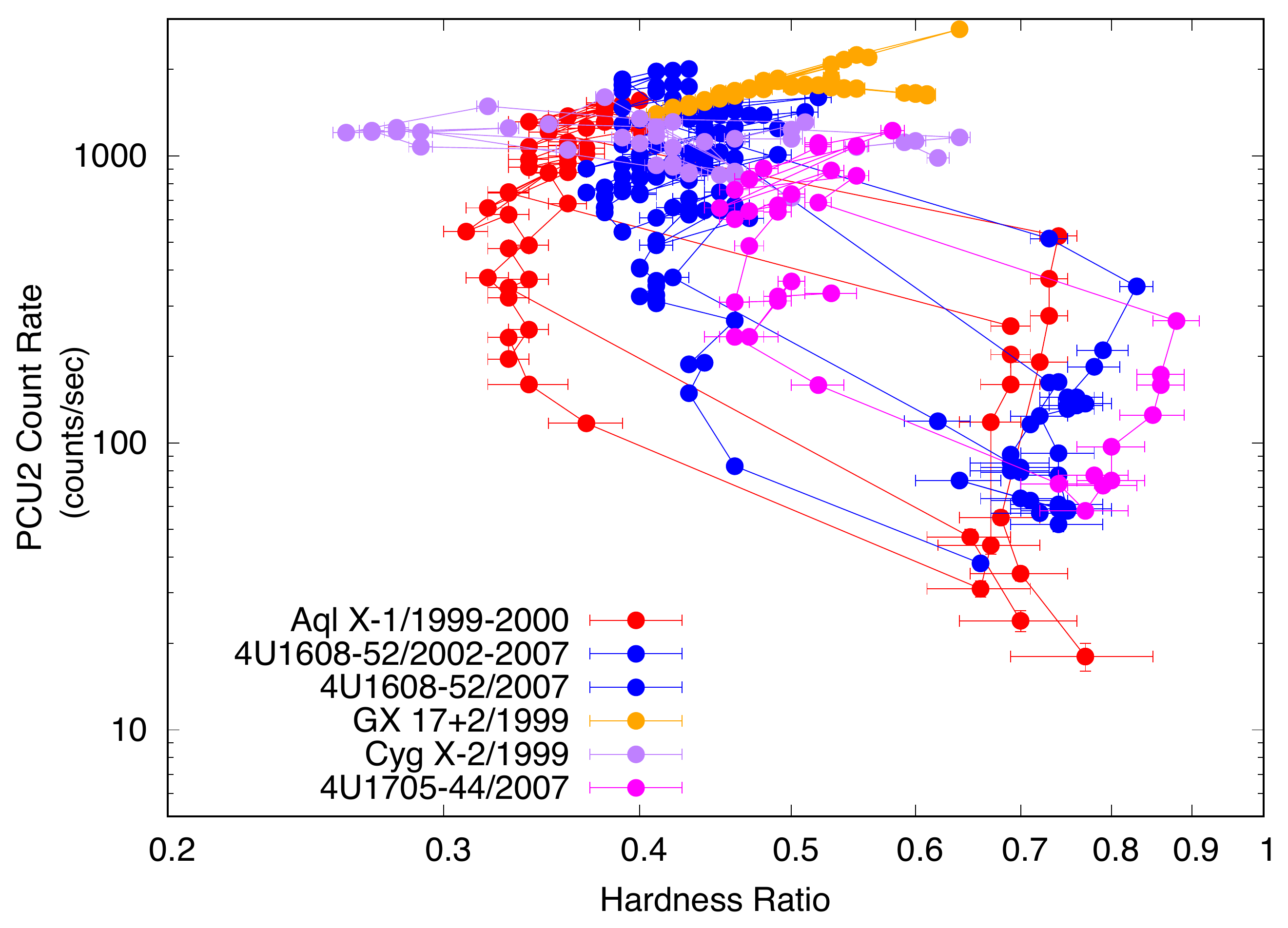}   
	\caption{Count rate (PCU2) vs Hardness ratio for the NS sources showing several hysteresis loops: hard and soft states are well separated.}   
	\label{fig1}
\end{figure}

\begin{figure}
	\hspace{-1.1cm}
	\centering      
	\includegraphics[scale=0.43, angle = 0]{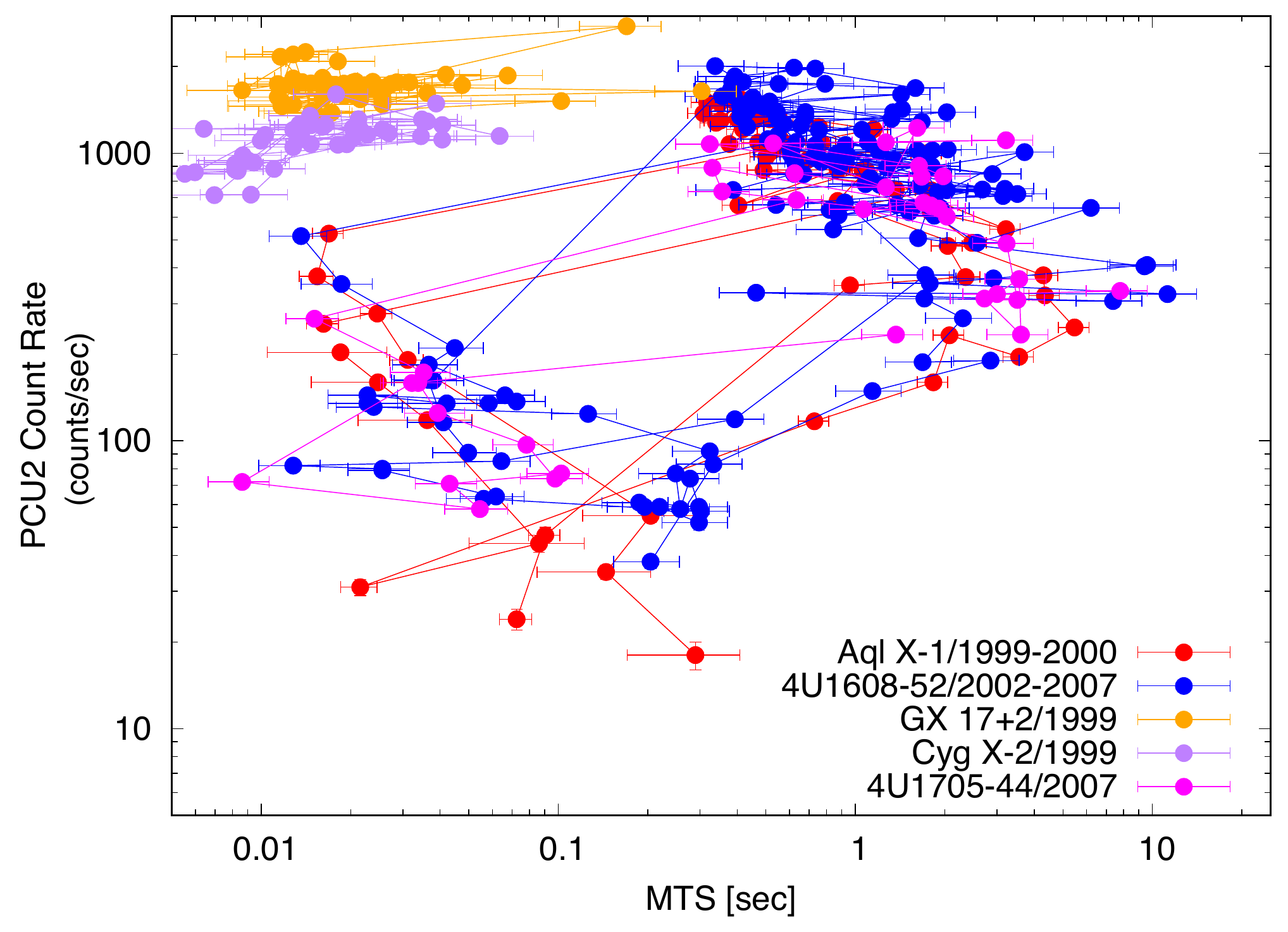}   
	\caption{Intensity-variability diagram (IVD) i.e., count rate (PCU2) vs MTS for the NS sources showing several hysteresis loops: hard and soft states are well separated. The IVD appears to be a 'mirror' image  of the equivalent HID (see Figure \ref{fig1}) but note the presence of a separate cluster due to Z-type sources at very small MTS.}    
	\label{fig2}
\end{figure}

\begin{figure}
	\hspace{-1.1cm}
	\centering      
	\includegraphics[scale=0.43, angle = 0]{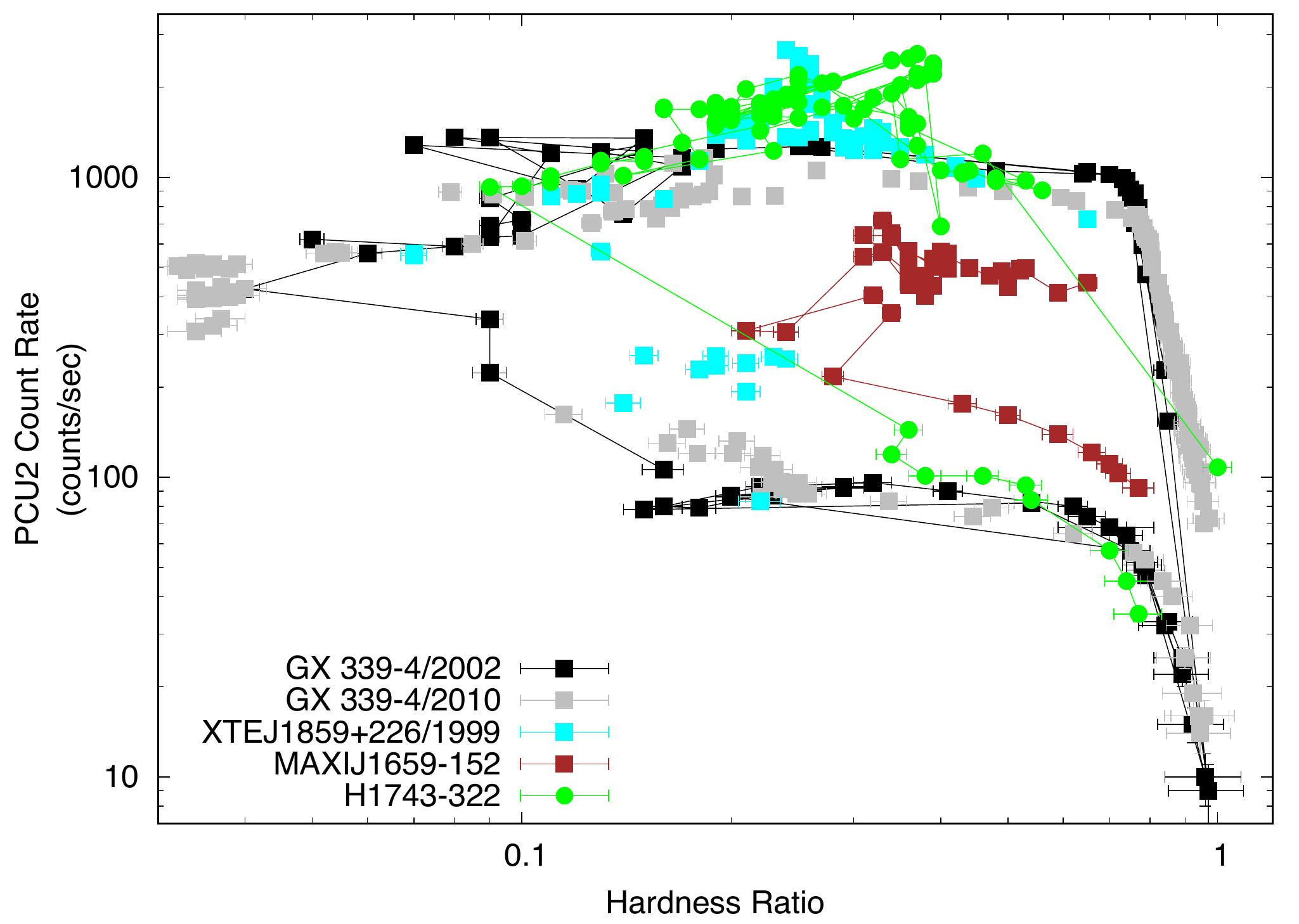}   
	\caption{Count rate (PCU2) vs Hardness for the BH sources showing transitions during several outbursts. See text for details.}    
	\label{fig3}
\end{figure}

\begin{figure}
	\centering      
	\includegraphics[scale=0.43, angle = 0]{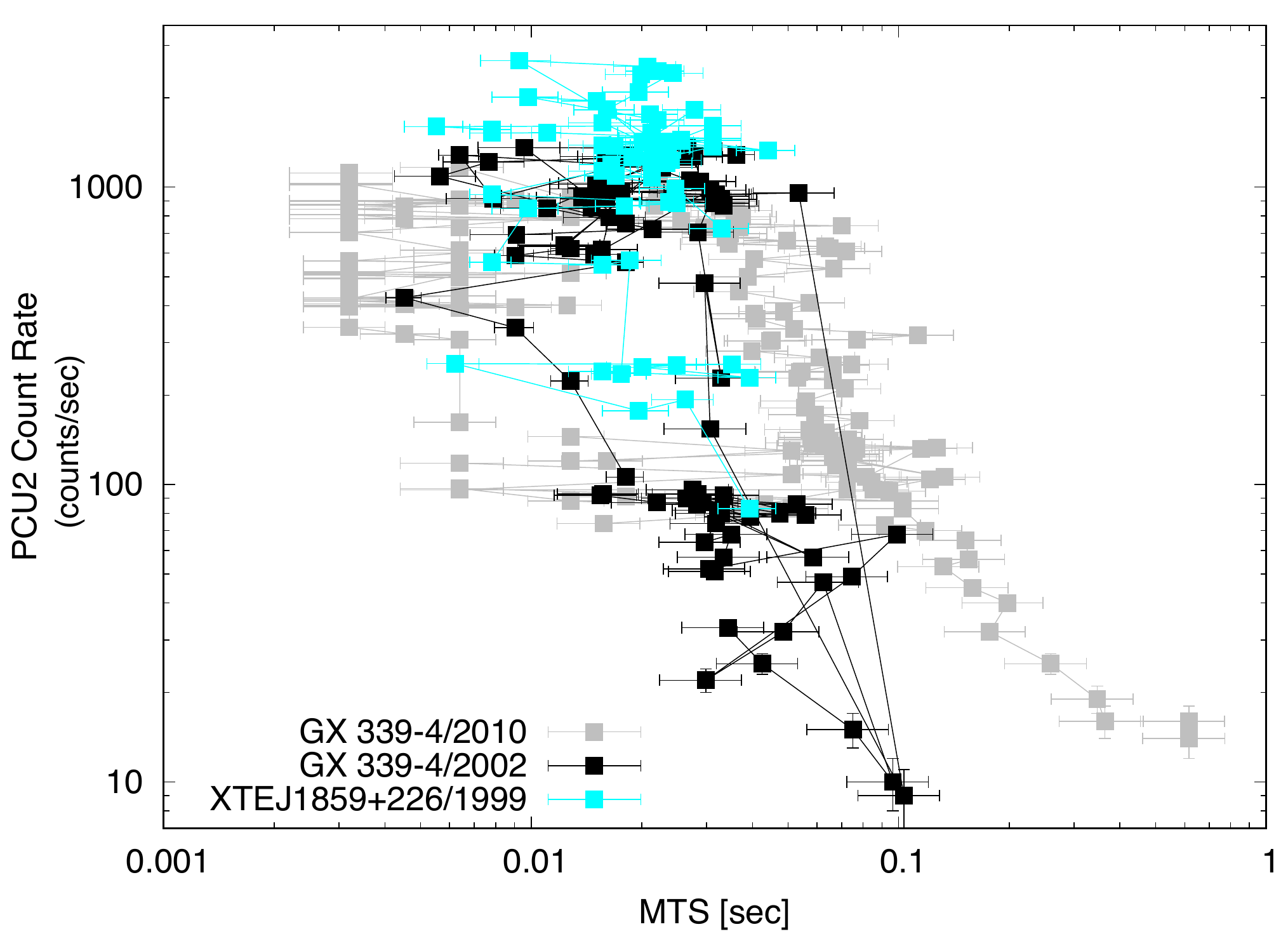}   
	\includegraphics[scale=0.43, angle = 0]{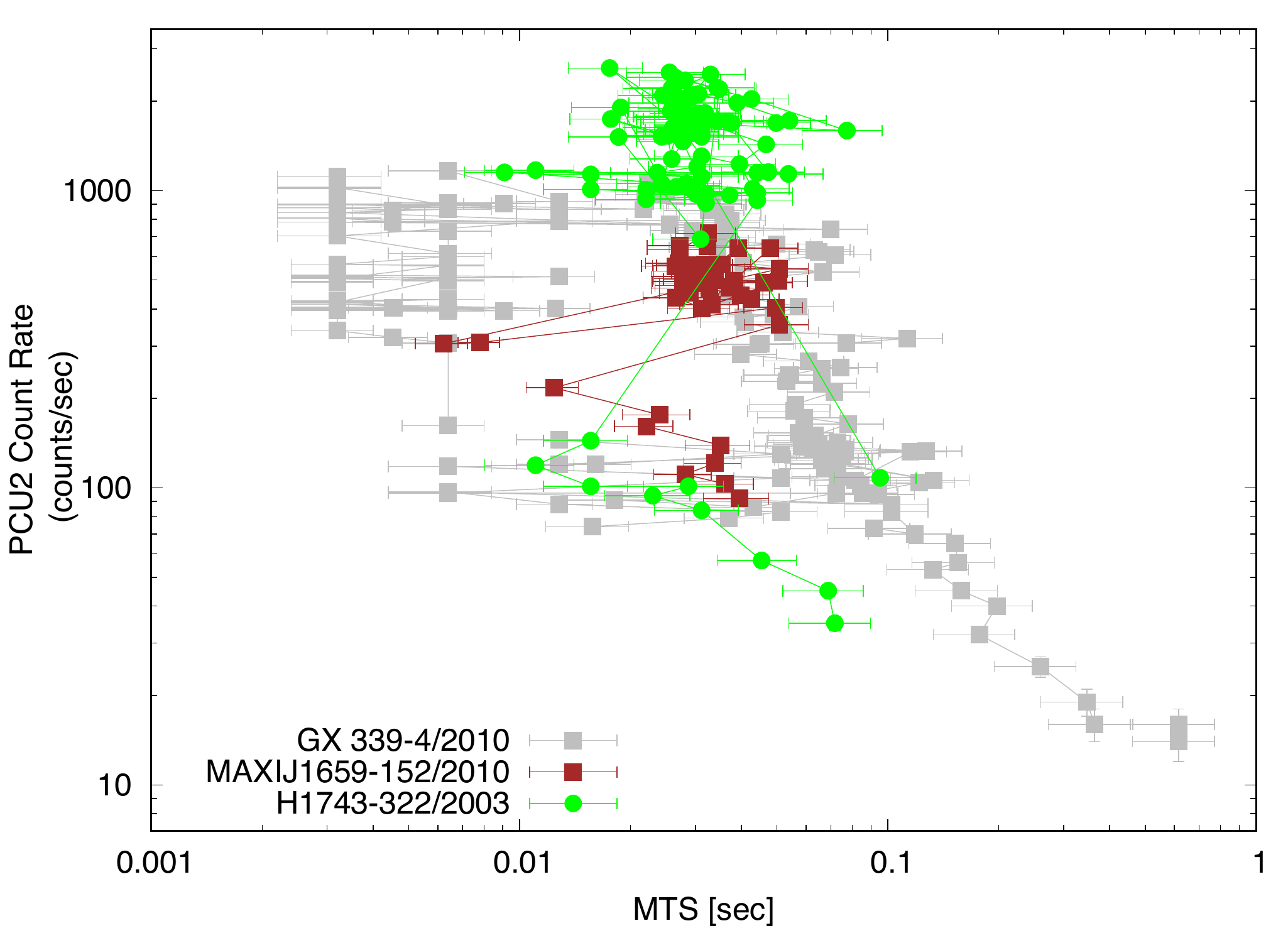}
	\caption{Intensity-variability diagram (IVD) i.e., count rate (PCU2) vs MTS for the BH sources showing several loops; The tracks exhibit a similar pattern to those seen in the equivalent HID (see Figure \ref{fig3}). Upper panel: 2002-2003 and 2010 outburst  of GX 339-4 and 1999 outburst for XTE J1859+226. Bottom panel: 2010 outburst  of GX 339-4 and MAXI J1659-152 and 2003 outburst of H1743-322.}    
	\label{fig4}
\end{figure}

\begin{figure}
	\hspace{-1.1cm}
	\centering      
	\includegraphics[scale=0.65, angle = 0]{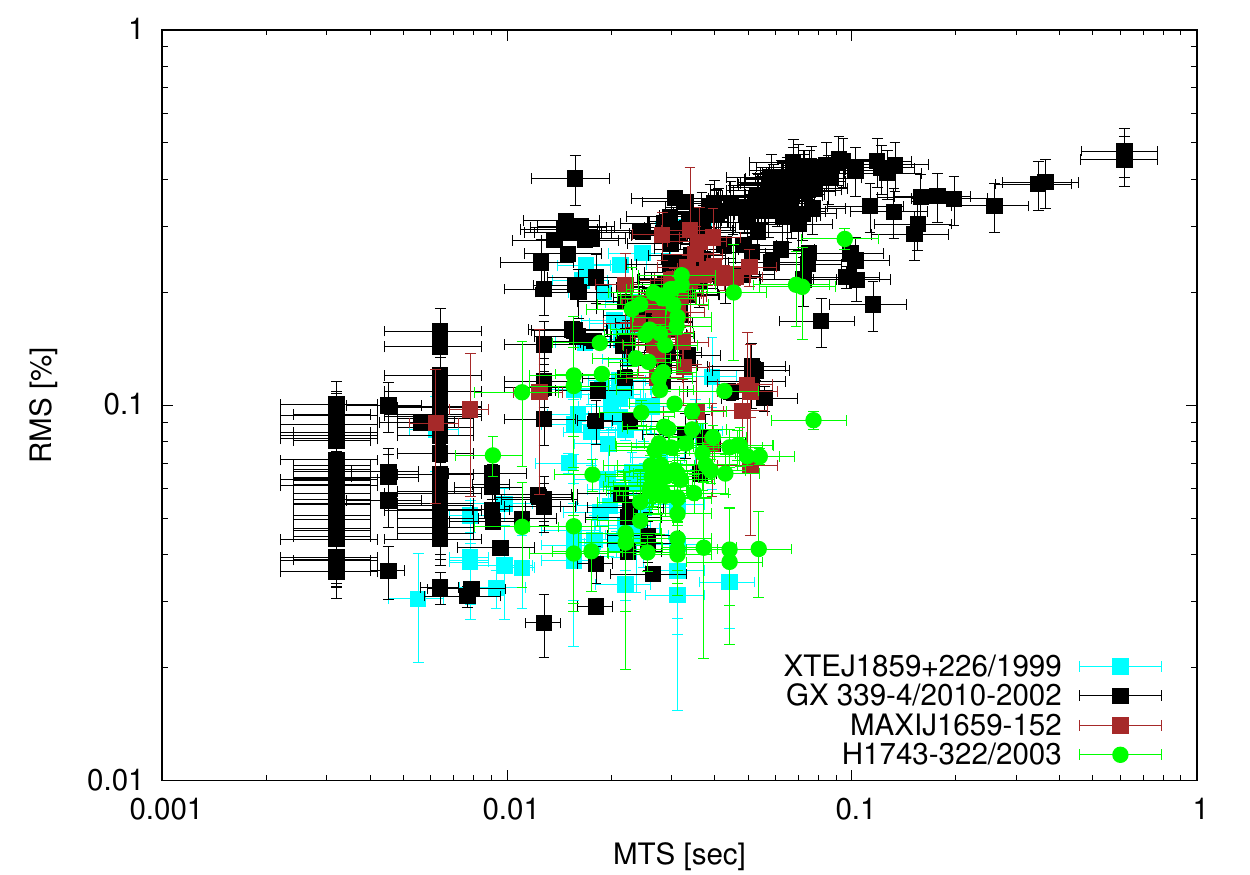}   
	\caption{Fractional RMS vs. MTS for several outburst of the BH transients GX339-4,  XTE J1859+226, MAXI J1659-152 and H 1743-322. The positive trend is clear.}    
	\label{fig5}
\end{figure}

\begin{figure}
	\hspace{-1.1cm}
	\centering      
	\includegraphics[scale=0.65, angle = 0]{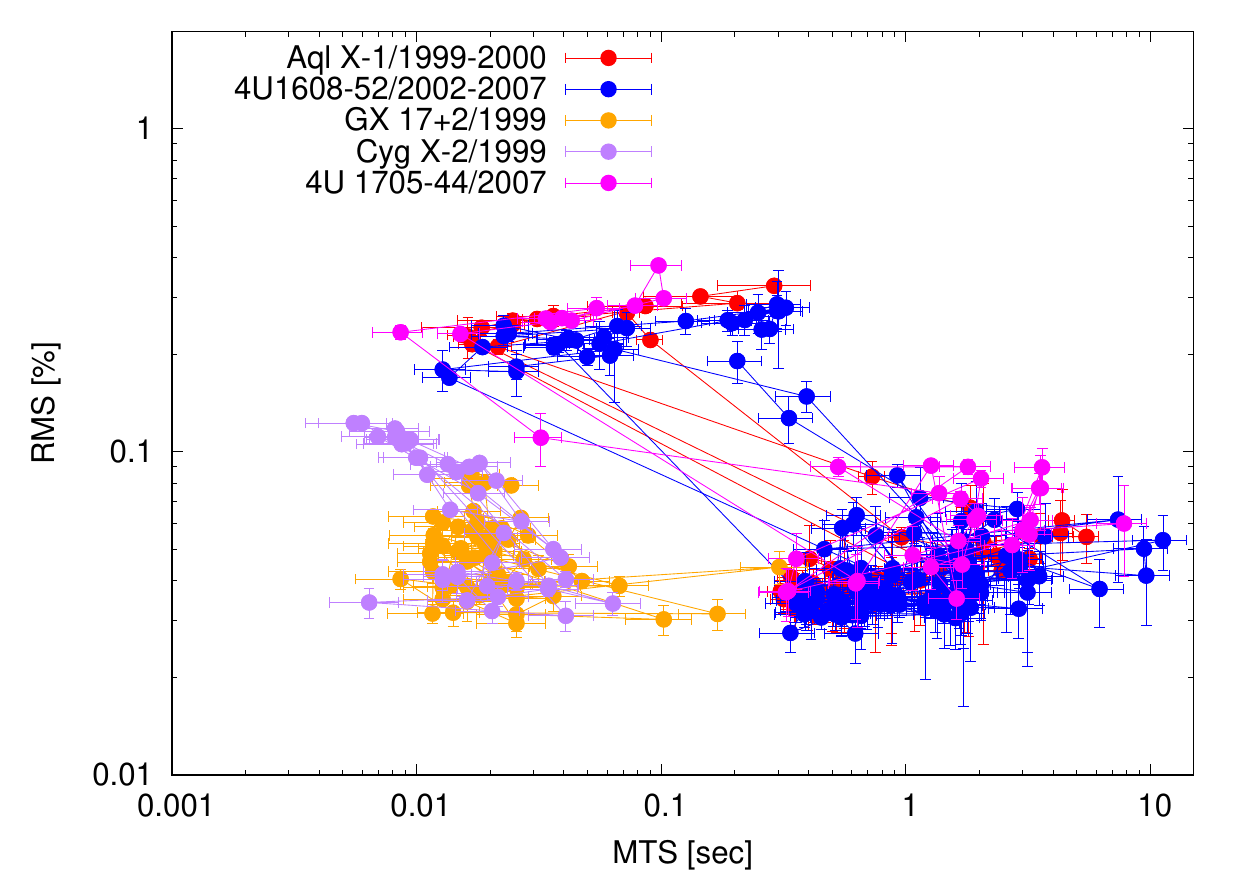}   
	\caption{Fractional RMS vs MTS for several loops of the NS-LMXBs: hard and soft states are well separated. Note the negative trend in contrast to the case of the BH transients shown in Figure 5. Furthermore,  the separation of the Z-type sources at small MTS is clearly evident.}   
	\label{fig6}
\end{figure}

\begin{figure}
	\hspace{-1.1cm}
	\centering      
	\includegraphics[scale=0.65, angle = 0]{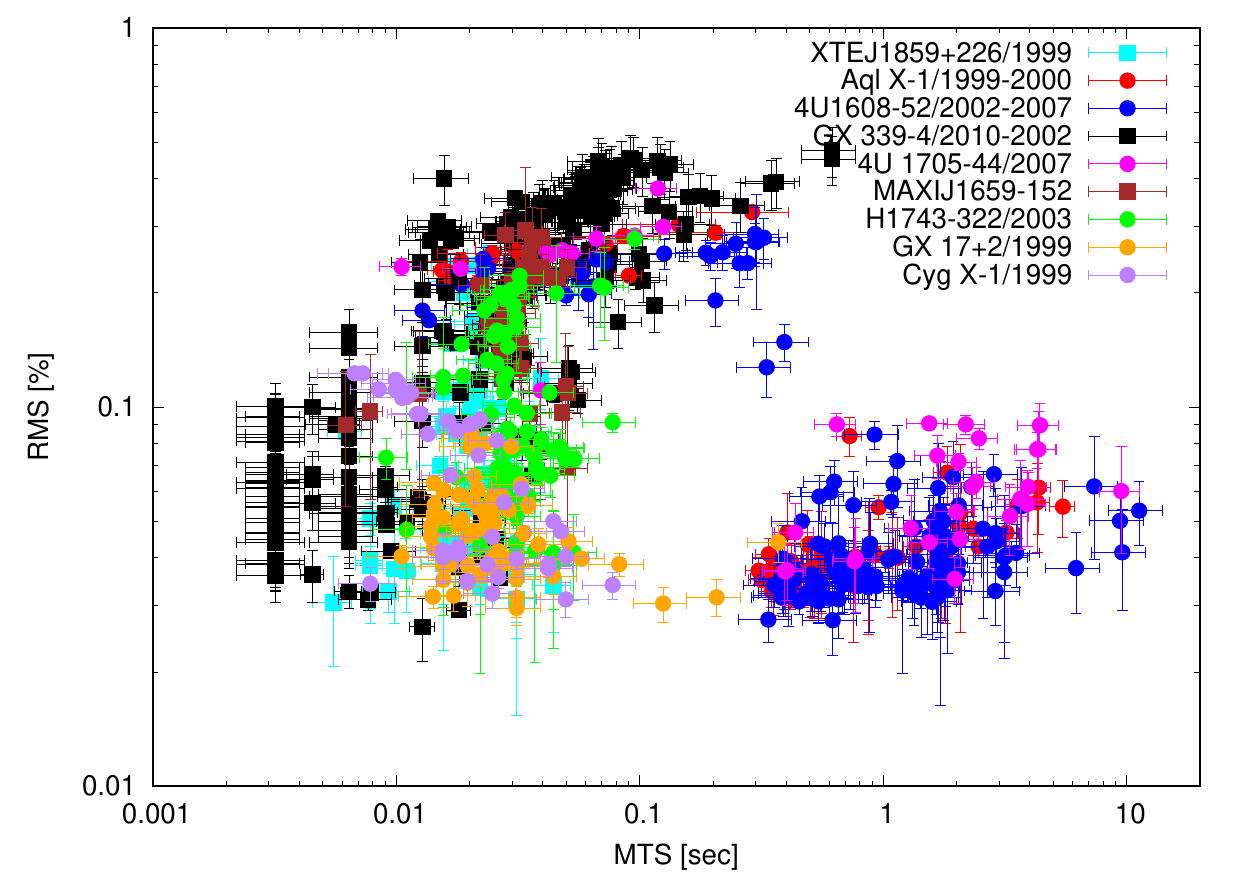}   
	\caption{Fractional RMS vs. MTS for the test sources.  High fractional RMS observations cluster in one common region occupied jointly by BH and NS sources; the low fractional RMS observations are divided into two distinct regions, well separated by MTS.}    
	\label{fig7}
\end{figure}

\begin{figure}
	\hspace{-1.1cm}
	\centering      
	\includegraphics[scale=0.65, angle = 0]{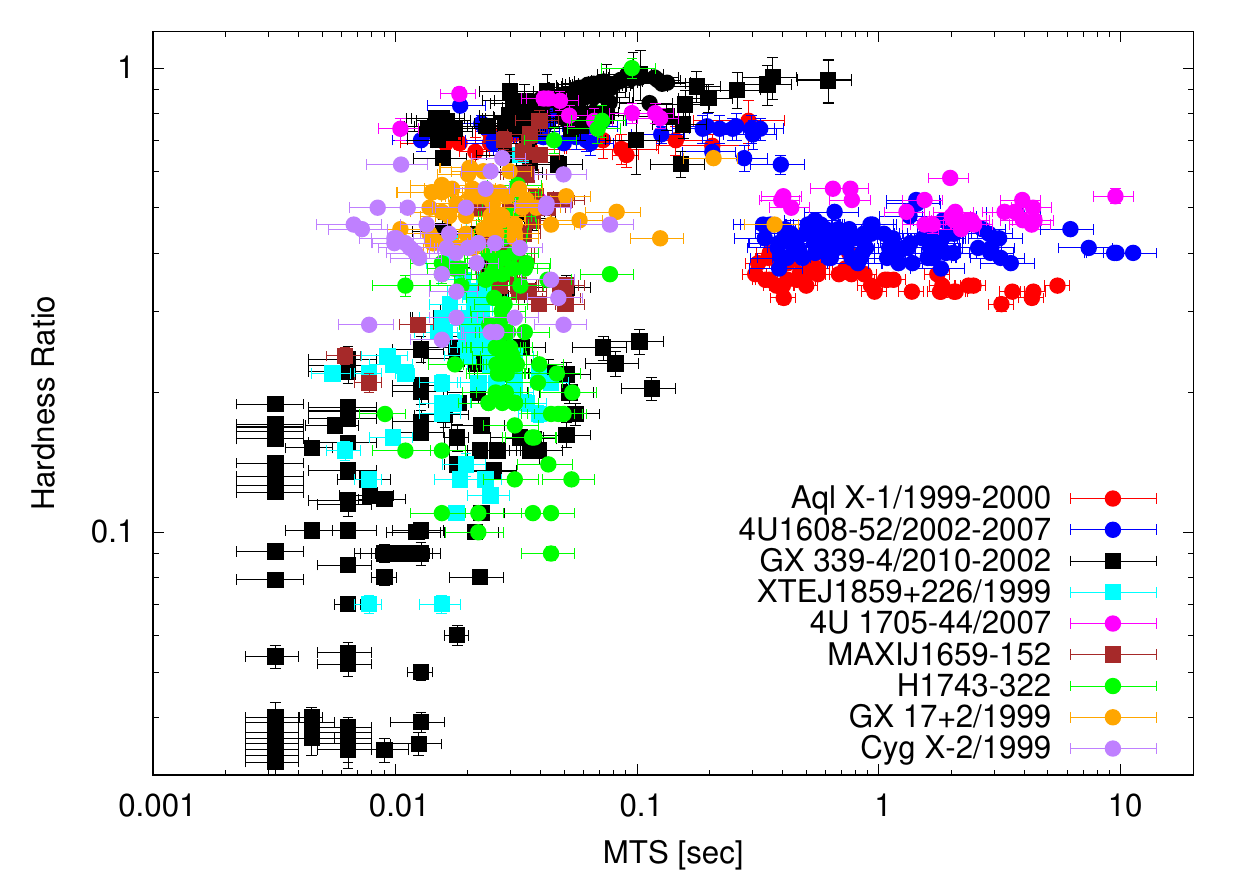}   
	\caption{Hardness ratio vs MTS for the test sources. BH and NS sources jointly occupy one common region corresponding to the (BH) low/hard state. However, the sources split into different regions for the soft and intermediate states with some separation in MTS. Note the absence of NS observations in the small MTS and very-low hardness region.} 
	\label{fig8}
\end{figure}

\begin{figure}
	\hspace{-1.1cm}
	\centering      
	\includegraphics[scale=0.25, angle = 0]{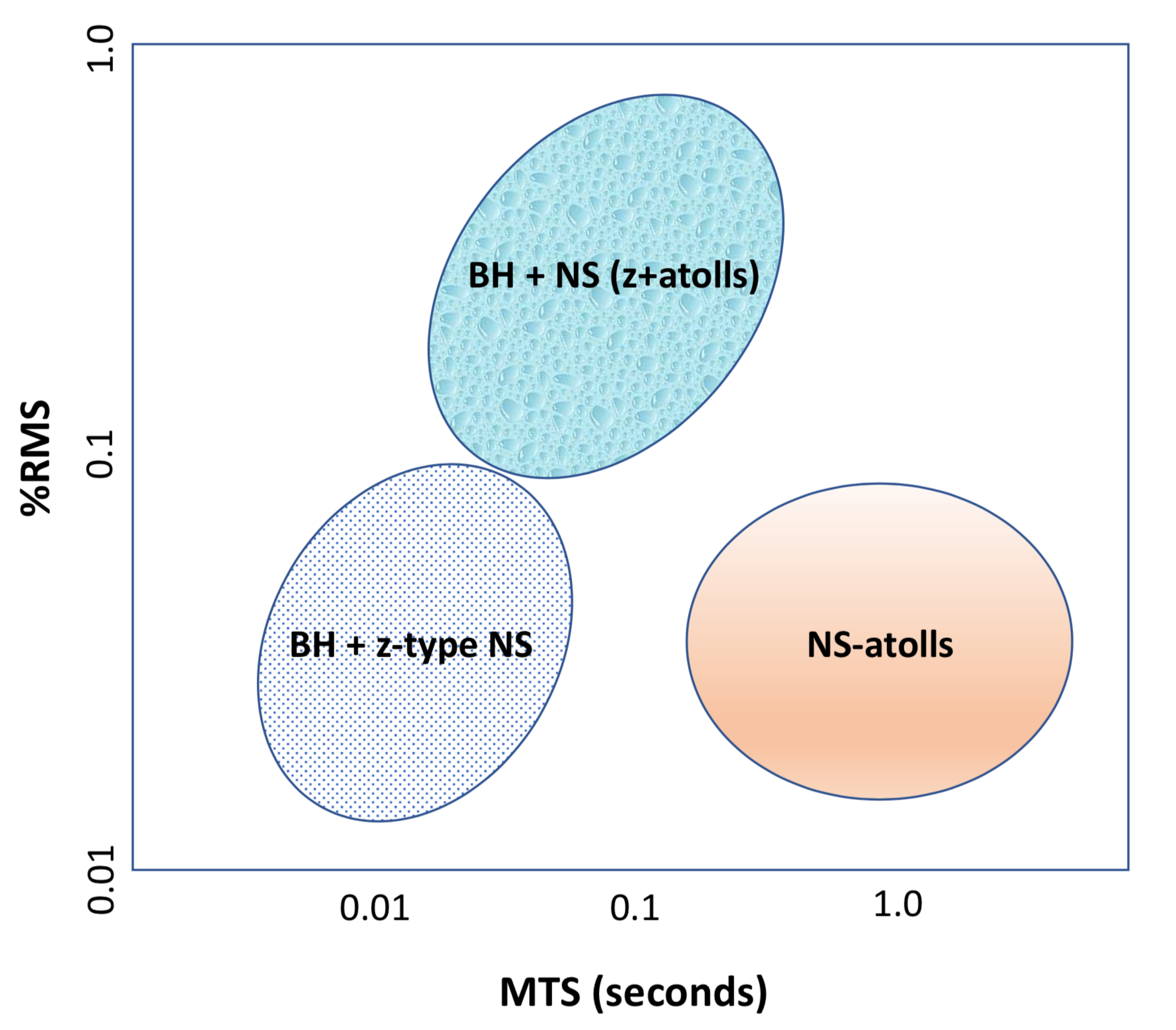}   
	\caption{A schematic representing the fractional RMS vs. MTS for a sample of BH and NS sources.  For high RMS, all sources occupy a  broad common region  in the upper portion of the plot. For low RMS, the Atolls and Z-type sources are well separated in MTS. BHs and Z-types share the low RMS and the small MTS region in the lower portion of the plot.}    
	\label{fig9}
\end{figure}

\begin{figure}
	\hspace{-1.1cm}
	\centering      
	\includegraphics[scale=0.25, angle = 0]{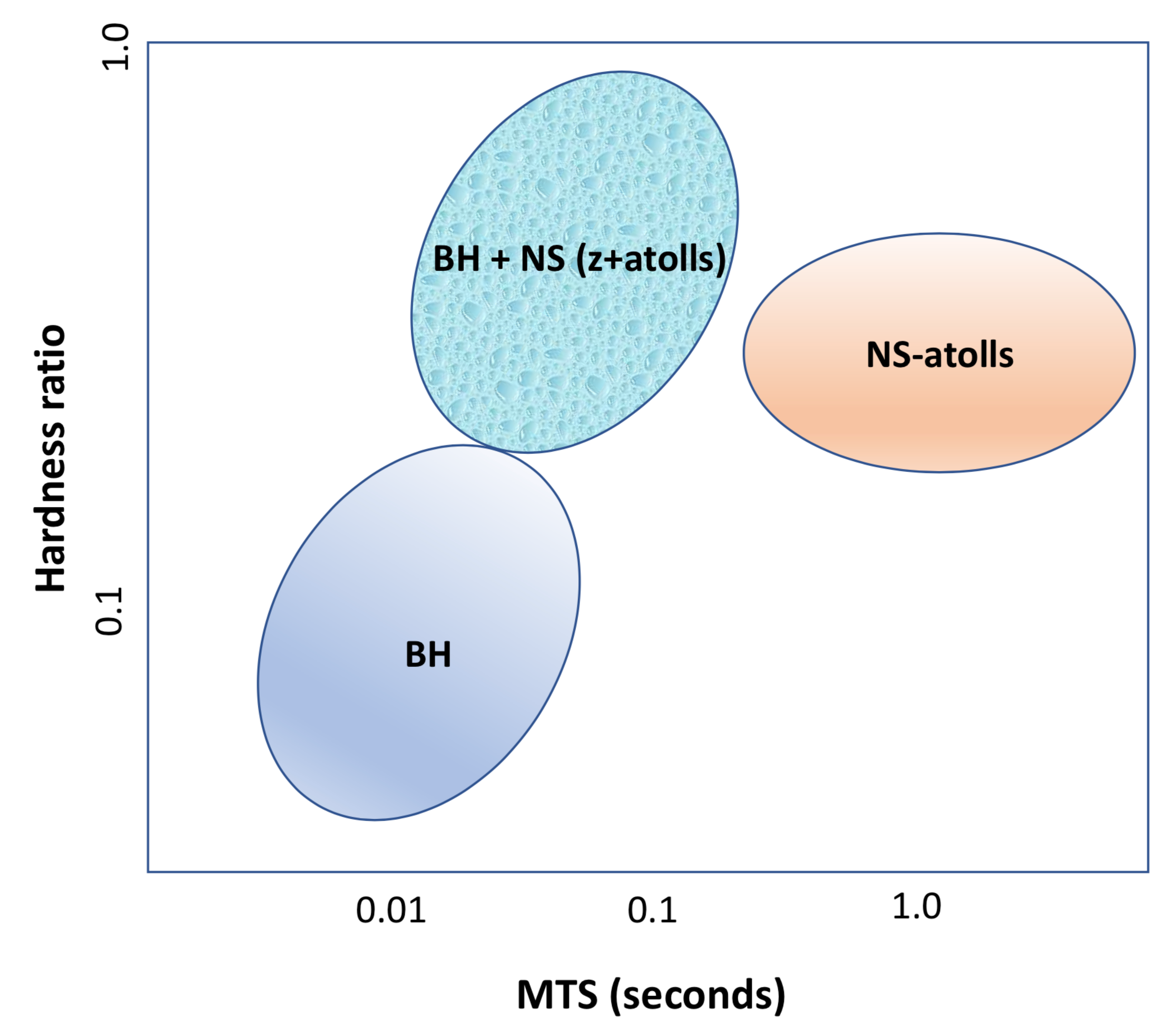}   
	\caption{A schematic representing the Hardness ratio vs MTS for a sample of BH and NS sources. For high hardness, all sources occupy a  broad common region  in the upper portion of the plot. The Atolls occupy a separate region at large MTS and relatively mid-range hardness. BHs are the sole occupiers of the very low hardness and small MTS region.}    
	\label{fig10}
\end{figure}

\noindent
Following the work of \cite{2014MNRAS.443.3270M}, who deployed both hardness and the RMS - intensity diagrams in their study of fast aperiodic variability of a sample of NS-LMXBs, we display in Figure \ref{fig1}, the HID for 4U1608-52, Aql X-1, 4U1705-44, GX17+2, and Cyg X-2. The plot clearly exhibits hysteresis-like loops similar to those observed in the sample studied by \cite{2014MNRAS.443.3270M}. The hardness (the ratio of counts within the 6-15 keV and 2-6 keV bands),  along with the RMS, and the extracted MTS values, are listed in Tables 1-2. For the same set of observations, we also constructed the IVDs i.e., the intensity (count/rate) - MTS plots over the complete range of transitions (see Figure \ref{fig2}). We see a clustering of the majority of the observations (for the Atoll sources) in two dominant spectral states, with a clear separation in MTS. Moreover, in the same figure, we notice an additional cluster of points at small MTS and relatively high count rates: these observations are from the Z-type sources (GX17+2 and Cyg X-2):  i.e., there appears to be a separation of Atolls and Z-type sources in the count-rate/MTS plane. \\
\\
For comparison, we show the same results (see Table 3) for the BH sources (i.e., GX339-4, XTE J1859+226, H1743-322, and MAXI J1659-152) in Figures \ref{fig3} and \ref{fig4}. In Figure \ref{fig3}, we see that GX339-4 (for both the 2002 and 2010 outbursts) and XTE J1859+226 (for the 1999 outburst) display q-shape trajectories as the sources transition across their states. The sources H1743-322 and MAXI J1659-152, with observations analyzed in this work, exhibit partial q-shape patterns.\\
\\ 
The NS sources show similar HID tracks (Figure \ref{fig1}) but exhibit noticeable differences between the relative intensity of the soft and hard states. Furthermore, the range of the hardness ratio is considerably smaller with lower level of $\sim$ 0.3 compared to $\sim$ 0.1 for the BH sources. However, the dramatic difference between the NS and BH sources is seen in Figures \ref{fig2} and \ref{fig4} where we have plotted the count rate vs MTS (i.e, the IVDs): The IVD for the BH sources (Figure \ref{fig4}) mimics the q-shape tracks seen in the corresponding HID (Figure \ref{fig3}). Whereas the combined IVD for the NS sources: 2 Atolls and 3 Z-types (Figure \ref{fig2}), is almost the mirror image of its corresponding HID (shown in Figure \ref{fig1}). Indeed, the soft state (a hardness ratio of $\sim 0.3$; Figure \ref{fig1}) exhibits the larger MTS (Figure \ref{fig2}), and it is the hard state (a hardness ratio of $\sim$ 0.7; Figure \ref{fig1}) that corresponds to the smaller MTS (Figure \ref{fig2}). Again, notice the separate cluster of points (in the top left corner of Figure \ref{fig2}) for the Z-types exhibiting a relatively small MTS. The behavior of the IVD in the case of the NS sources (at least the Atolls) appears to be in complete contrast to the case of the BH sources where the MTS appears to act as a proxy for the hardness i.e, the soft state corresponds to the smaller MTS and the hard state corresponds to the larger MTS (Figure \ref{fig4}). The Z-types, on the other hand, seem to be a lot more similar to the BH sample in this regard i.e, smaller MTS corresponds to smaller hardness. We revisit this particular point of apparent dichotomy between NS- and BH-LMXBs later in the discussion.
\\
\\
A traditional measure of the variability is the RMS which is normalized with respect to the count rate and is typically extracted from a suitably normalized PSD (and quoted as a percentage). \cite{2020MNRAS.853.150S} in their study of GX339-4, investigated a possible connection between RMS and MTS and reported a positive correlation between them, albeit with large scatter. They also noted that the correlation is likely non-causal, hinting instead that the individual variables were more likely exhibiting an independent sensitivity to (yet) an unknown common physical parameter pertinent to the accretion process itself. We have reproduced the analysis of \cite{2020MNRAS.853.150S} for the BH transient GX339-4, and in addition, report results of a similar analysis for XTE J1859+226, H1743-322 and MAXI J1659-152. The results, presented in Figure \ref{fig5}, confirm the positive correlation between RMS and MTS. At this juncture, we are in a position to make a direct comparison of the RMS-MTS response of the BH and NS sources; see Figures \ref{fig5} and \ref{fig6} respectively. The results indicate a correlation but in the case of the Atoll NS sources (Figure \ref{fig6}) the trend is clearly an anti-correlation, in agreement with the results reported by \citet{2021arXiv210104201M}. The Z-types form a separate cluster, shifted to a lower MTS, with a hint of an anti-correlation as well.\\
\\
In order to further probe the apparent dichotomy between the NS- and BH-LMXBs, we show the combined results of RMS, MTS and hardness ratios for our test sources in Figures \ref{fig7} and \ref{fig8}. It is quite clear from these figures that the NS and BH sources separate themselves into different sections of the RMS-MTS and Hardness-MTS planes: a fairly broad common region, occupied by both BH and NS sources, and several separate regions: one jointly occupied by BHs and Z-type NS sources; two regions solely occupied by Atolls, and BHs respectively. For a better visualization of these results, we present the various regions in two schematics representing the RMS-MTS (Figure \ref{fig9}) and Hardness-MTS (Figure \ref{fig10}) planes respectively. The positions of the respective sources in these schematics not only allow for easy visualization but also sets the stage for interpreting the results. First, we make a few general comments about the two schematics; they identify three regions in the relevant parameter space, where each region is labeled according to the type of source that occupies that region. The oval shapes are used to indicate clustering and  do not imply strict boundaries. The scales are also representative of our sample and are likely to be different for other sources. More specifically, we note the following from the RMS-MTS schematic: For low RMS, MTS (on average) cleanly separates Atolls and Z-type sources i.e., small MTS corresponds to Z-types and large MTS corresponds to Atolls. 
One possibility that comes to mind is related to the accretion rate;  in Atolls the luminosity is a small fraction of Eddington typically 10$\%$ whereas in Z-type sources the luminosity is significantly higher i.e., typically greater than 50$\%$ of Eddington. With other parameters being similar, the large difference in luminosity is likely to be a reflection of the accretion rate in the respective types of sources i.e., the accretion rate is expected to be larger at least by a factor of five in Z-type sources. The overall impact is potentially a larger coronal density in Z-type sources (an optically thick accretion flow) compared to the situation in Atolls where the accretion rate is expected to be much lower and therefore a lower coronal density (an optically thin accretion flow). In turn, this implies larger interaction distances and longer time scales in the case of Atoll sources. Ostensibly, another factor that one might consider is the magnetic field strength in Atolls and Z-type sources. A sufficiently strong field (of the order 10$^{10}$ G) can re-direct the accretion flow to the polar regions effectively truncating the accretion disk thus significantly altering the geometry of the accretion and the availability of seed photons. However, this is unlikely to be the main driving term for the difference in the observed MTS between the two types of sources as the mean field strength of the Atolls and Z-type sources in our sample is around 10$^8$ G \citep{2010ApJ...719.1350L}. \\
\\
We see from the RMS-MTS schematic that BHs too can occupy the region of low RMS and small MTS; this is the case of BHs in the high/soft state, where the accretion disk, in principle, can reach into the vicinity of the ISCO, and thus accretion in BHs provides the full spectrum of timescales from the very small to the large (when the disk is truncated as is believed to be the case for the low/hard state). Interestingly, we see small (MTS) timescales in Z-type sources as well. This may hint at the existence of an accretion-flow environment similar to that assumed to exist in the case of BHs in the high/soft state where the interaction is presumably dominated by the inner region of the non-truncated accretion disk.\\
\\
The Hardness-MTS schematic, depicted as Figure \ref{fig10}, is also very informative. The figure shows that the region corresponding to the very low-hardness/low-MTS values is occupied solely by BH sources. Apparently, NS sources do not reach this very low hardness (a feature attributed to the presence  of the hard surface in NS sources, see \cite{2003MNRAS.342..361N}). We noted above that Z-type sources may exhibit a similar emission environment to that thought to exist for BHs in the high/soft state. The environments may be similar (coronal density for example) but cannot be identical in the two cases because of the existence of a surface and a sizable magnetic field in the case of NS sources. Indeed, the BHs and the Z-type sources do share a region in Figure \ref{fig10} but clearly the hardness scale separates the BH and NS sources such that the very low hardness and low MTS part is solely occupied by BHs.\\
\\
The combined findings represented in Figures \ref{fig7} and \ref{fig8} (represented by schematics \ref{fig9} and \ref{fig10} respectively) are in excellent agreement with the work of \cite{2003MNRAS.342..361N} (and see the more recent studies of \citet{2018MNRAS.481.3761G} who used power color-color diagrams). \cite{2003MNRAS.342..361N} studied a sizable sample of BH and NS LMXBs and demonstrated a clear dichotomy between the NS and BH sources by showing that the sources occupy distinct regions in a color-color diagram. In particular, they noted that BH sources reach the high/soft state, depicted by a very low hardness ratio whereas the NS sources, by comparison, only reach moderate hardness, a difference they attributed to the presence of the hard surface. 
\section{Summary and Conclusions}
\noindent
 In this study, we analyzed publicly available RXTE/PCA observations of 9 LMXBs: 5 that host a NS primary (4U1608-52, Aql X-1, 4U1705-44, GX17+2 and Cyg X-2), and 4 that host a BH (GX339-4, XTE J1859+226, H1743-322 and MAXI J1659-152). The NS group includes 3 Atolls and 2 Z-type sources. For all of the observed spectral states, we constructed normalized PSDs and extracted the fractional RMS variability by direct integration.  We extracted a minimal time-scale (MTS) from the measured lightcurves by using fast wavelet transforms. Through a comparison of the extracted RMS, the hardness ratios, and the MTS results, we have probed the dichotomy between BH and NS binary sources. Our main findings are summarized as follows;\\   
\begin{itemize}
\item We use the extracted MTS to construct an intensity-variability diagram (IVD) for all the outbursts analyzed for each source. For the BH sources the IVD exhibits hysteresis-like loops similar to the q-shape tracks seen in the standard HID plots. \\
\item In contrast, the IVD for the NS (Atoll) sources appears to be a 'mirror' image of the equivalent HID. The Z-type sources form a separate cluster in the low-MTS region.\\
\item We confirm the results of \cite{2020MNRAS.853.150S} and \cite{2021arXiv210104201M} by linking the extracted RMS variabilities to the characteristic time scales (MTS) for each lightcurve and demonstrate (albeit with considerable scatter) a positive correlation for BH transients and an anticorrelation for the Atoll NS sources. In this  regard, the evidence for Z-type sources, which form a separate cluster at small  MTS, is less compelling.\\
\item The IVD, the RMS-MTS and the hardness-MTS plots, all indicate that the NS transitions are dominated by two main spectral states, a hard state and a relative soft state. The NS hard state occupies a similar region to that occupied by BH transients in all of the diagrams thus suggesting the equivalency of this state with the low/hard state of BH sources.\\
\item In contrast to the hard state, the relative soft state in NS sources occupies a distinctly different position in both the RMS-MTS and the hardness-MTS plots, suggesting a different spatial origin (and/or process) for the emission associated with this state.\\
\item For low RMS observations, the MTS (on average) separates Atolls from Z-type and BH sources: small MTS corresponds to Z-types (and  BHs), and large MTS to Atoll sources.\\ 
\item In the hardness-MTS plane, BH sources appear to be the sole occupiers of the very low hardness-small-MTS region. This unique feature has the potential to discriminate between BH and NS hosts in LMXBs. The absence of observed emission in this region for NS sources is likely due to the presence of the hard surface in these sources.\\
\end{itemize}
\section*{Acknowledgements}
\noindent
This research is supported by the Scientific and Technological Research Council of Turkey (TUBITAK) through project  number 117F334. In addition, KM acknowledges financial support provided by the Cultural Affairs and Missions Sector Ministry of Higher Education of Egypt. \\

\section*{Data availability}
 The RXTE (Rossi X-ray Timing Explorer) data analyzed in the course of this study are available in HEASARC, the  NASA's Archive of Data on Energetic Phenomena. The full versions of the data tables underlying this article are available as online supplementary material.
\begin{table*}
	\centering
	\caption{RMS and MTS for NS (Atoll) LMXBs. Full version of the table is available online.}
	\label{tab:example_table}
	\begin{tabular}{rrrrrrrrr} 
	\hline
Obs. Id. & RMS &   $\tau$   & Hardness & PCU2 \\
	\hline
& $\%$ &  sec & & cnts/sec \\
\hline\hline
{\bf 4U1608-52/2002}  \\
\hline
70058-01-10-00	& 	19.7	$\pm$	1.2	& 	0.050	$\pm$ 	0.012	&0.69 $\pm$ 0.04&91$\pm$3	\\
70058-01-11-00	& 	22.2$\pm$ 	1.6	& 	0.041	$\pm$ 	0.010& 0.71$\pm$ 0.03&116$\pm$4	\\
70058-01-12-00	& 	21.0$\pm$		0.9	& 	0.037	$\pm$	0.009& 0.74$\pm$ 0.03&163$\pm$4	\\
70058-01-13-00	& 	21.6$\pm$ 	1.3	& 	0.038	$\pm$ 	0.010& 0.73 $\pm$ 0.03&162$\pm$	4\\
70059-01-01-03	& 	3.5	$\pm$ 	0.5	& 	0.392	$\pm$ 	0.098& 0.39 $\pm$ 0.01&1850$\pm$11	\\
70059-01-01-02	& 	3.5	$\pm$ 	0.6	& 	0.732	$\pm$ 	0.183& 0.41 $\pm$ 0.01&1968$\pm$	12\\
70058-01-15-00	& 	2.8	$\pm$ 	0.4	& 	0.337	$\pm$ 	0.084&0.43 $\pm$ 0.01&2006$\pm$	12\\
70058-01-16-00	& 	3.7	$\pm$ 	0.4	& 	0.418	$\pm$ 	0.104& 0.39$\pm$ 0.01&1767$\pm$11	\\
70058-01-17-00	& 	2.7	$\pm$ 	0.5	& 	0.621		$\pm$ 	0.155	&0.42 $\pm$ 0.01&1983$\pm$12	\\
70058-01-19-01	&	3.1	$\pm$	0.6	&	1.594	$\pm$	0.399&0.39$\pm$0.01&1685$\pm$11	\\
\hline\hline
{\bf 4U1608-52/2007}  \\
\hline
93408-01-11-02&	27.28$\pm$	3.19&	0.298$\pm$	0.074&	0.74$\pm$	0.05&	52$\pm$	3\\
93408-01-11-04&	25.59$\pm$	2.73&	0.219$\pm$	0.055&	0.74$\pm$	0.05&	59$\pm$	3\\
93408-01-11-03&	25.02$\pm$	2.07&	0.195$\pm$	0.049&	0.75$\pm$	0.05&	59$\pm$	3\\
93408-01-12-03&	25.55$\pm$	1.91&		0.187	$\pm$	0.047&	0.74$\pm$	0.05&	61$\pm$	3\\
93408-01-12-00&	28.60$\pm$	5.08&	0.298$\pm$	0.074&	0.74$\pm$	0.05&	59$\pm$	3\\
93408-01-12-01&	23.99$\pm$	2.34&	0.278$\pm$	0.070&	0.64$\pm$	0.04&	74$\pm$	3\\
93408-01-12-02&	23.92$\pm$	3.23&	0.257$\pm$	0.065&	0.75$\pm$	0.05&	58$\pm$	3\\
93408-01-13-01&	26.97$\pm$	3.75&	0.249$\pm$	0.062&	0.74$\pm$	0.04&	77$\pm$	3\\
93408-01-13-02&	27.96$\pm$	3.42&	0.324$\pm$	0.081&	0.74$\pm$	0.04&	92$\pm$	3\\
93408-01-13-03&	25.39$\pm$	2.35&	0.126$\pm$	0.031&	0.72$\pm$	0.03&	124$\pm$  4\\
\hline\hline
{\bf Aql X-1/1999}  \\
\hline
40047-01-01-00	& 	22.8	$\pm$ 	3.2	& 	0.016	$\pm$ 	0.002& 0.69 $\pm$0.02	&255$\pm$5	\\
40047-01-01-02	& 	4.0	$\pm$ 	1.3	& 	1.809	$\pm$ 	0.213& 0.33 $\pm$0.01	&742 $\pm$8	\\
40047-02-01-00	& 	4.3	$\pm$ 	1.0	& 	0.491	$\pm$ 	0.058& 0.35   $\pm$   0.01  &874 $\pm$8 \\
40047-02-02-00	& 	3.6	$\pm$ 	1.1	& 	0.872	$\pm$ 	0.103&0.36 $\pm$	0.01&876$\pm$8	\\
40047-02-03-00	& 	4.7	$\pm$ 	1.2	& 	0.404	$\pm$ 	0.048&0.32 $\pm$0.01	&659$\pm$7	\\
40047-02-03-01	& 	4.0	$\pm$ 	1.2	& 	1.085	$\pm$ 	0.128&0.35 $\pm$0.01	&872	 $\pm$8\\
40047-02-04-00	& 	4.2	$\pm$ 	1.5	& 	0.874	$\pm$ 	0.103&0.36 $\pm$0.01	&682$\pm$7	\\
40047-02-05-00	& 	5.6	$\pm$ 	1.5	& 	4.307	$\pm$ 	0.508&0.32 $\pm$	0.01&376	$\pm$6\\
40047-03-01-00	& 	6.1	$\pm$ 	1.5	& 	4.357	$\pm$ 	0.514&0.33 $\pm$	0.01&320$\pm$5	\\
40047-03-02-00	& 	5.5	$\pm$ 	0.9	& 	5.468	$\pm$ 	0.644&0.34 $\pm$0.01	&248$\pm$	5\\
\hline\hline
{\bf Aql X-1/2000}  \\
\hline
50049-01-03-00&	32.7$\pm$	0.7&	0.290$\pm$	0.119&	0.77$\pm$	0.08&	18$\pm$		2 \\
50049-01-03-01&	30.3$\pm$	0.5&	0.145$\pm$	0.060&	0.70$\pm$	0.05&	35$\pm$		2\\
50049-01-03-02&	28.9$\pm$	1.2&	0.205$\pm$	0.084&	0.68$\pm$	0.04&	55$\pm$		3\\
50049-01-04-00&	25.8$\pm$	0.4&	0.031$\pm$	0.004&	0.72$\pm$	0.03&	191$\pm$		4\\
50049-01-04-01&	24.7$\pm$	0.6&	0.025$\pm$	0.003&	0.73$\pm$	0.02&	277$\pm$		5\\
50049-01-04-02&	23.0$\pm$	0.4&	0.015$\pm$	0.002&	0.73$\pm$	0.02&	373$\pm$		6\\
50049-01-04-03&	21.6$\pm$		1.1&	0.017	$\pm$	0.002&	0.74$\pm$	0.02&	526$\pm$		6\\
50049-01-04-04&	4.0$\pm$		1.1&	1.150$\pm$ 	0.135&	0.35$\pm$	0.01&	1210$\pm$	9\\
50049-01-05-00&	3.4$\pm$		0.4&	0.420$\pm$	0.050&	0.38$\pm$	0.01&	1418$\pm$	10\\
50049-01-05-01&	3.4$\pm$		0.4&	0.550$\pm$	0.065&	0.40$\pm$	0.01&	1337$\pm$	10\\
\hline\hline
{\bf 4U 1705-44/2007}  \\
\hline
93060-01-03-00&	7.45$\pm$		0.96&	1.670$\pm$	0.317&	0.46$\pm$	0.02&	234$\pm$ 5\\
93060-01-04-00&	11.06$\pm$	2.07&	0.039$\pm$	0.007&	0.52$\pm$	0.02&	159$\pm$	 4\\
93060-01-05-00&	23.44$\pm$	1.23&	0.011$\pm$	0.002&	0.74$\pm$	0.04&	72$\pm$	3\\
93060-01-06-00&	27.83$\pm$	2.13&	0.067$\pm$	0.013&	0.77$\pm$	0.05&	58$\pm$	3\\
93060-01-07-00&	25.47$\pm$	1.03&	0.053$\pm$	0.010&	0.79$\pm$	0.04&	71$\pm$	3\\
93060-01-08-00&	29.86$\pm$	1.30&	0.125	$\pm$	0.024&	0.78$\pm$	0.04&	77$\pm$	3\\
93060-01-09-00&	37.72	$\pm$	0.87&	0.119$\pm$	0.023&	0.80$\pm$	0.04&	74$\pm$	3\\
93060-01-10-00&	28.32$\pm$	0.83&	0.095$\pm$	0.018&	0.80$\pm$	0.04&	97$\pm$	3\\
93060-01-11-00&	25.94$\pm$	0.55&	0.048$\pm$	0.009&	0.85$\pm$	0.04&	125$\pm$	4\\
93060-01-12-00&	25.86$\pm$	0.44&	0.041$\pm$	0.008&	0.86$\pm$	0.03&	159$\pm$	4\\
\hline\hline
\end{tabular}
\end{table*}
\begin{table*}
	\centering
	\caption{RMS and MTS for NS (Z-type sources) LMXBs. Full version of the table is available online.}
	\label{tab:example_table}
	\begin{tabular}{rrrrrrrrr} 
	\hline
Obs. Id. & RMS &   $\tau$   & Hardness & PCU2 \\
	\hline
&  $\%$ &  sec & & cnts/sec \\
\hline\hline
{\bf GX 17+2/1999}  \\
\hline
40018-01-02-00&	6.25$\pm$		0.08&	0.033$\pm$		0.008&	0.550$\pm$		0.007&	1711$\pm$			11 \\
40018-01-02-08&	5.35$\pm$		0.09&	0.029$\pm$		0.007&	0.530$\pm$		0.007&	1772$\pm$		11\\
40018-01-02-20&	4.85$\pm$		0.49&	0.014$\pm$		0.003&	0.500$\pm$		0.006&	1739$\pm$		11\\
40018-01-02-21&	4.91$\pm$			0.28&	0.018$\pm$		0.004&	0.510$\pm$		0.006&	1750$\pm$		11\\
40018-01-02-18&	3.80$\pm$		0.37&	0.031$\pm$		0.008&	0.460$\pm$		0.006&	1684$\pm$		11\\
40018-01-02-19&	3.86$\pm$		0.28&	0.031$\pm$		0.008&	0.460$\pm$		0.006&	1689$\pm$		11\\
40018-01-02-01&	3.98$\pm$		0.14&	0.058$\pm$		0.014&	0.470$\pm$		0.006&	1725$\pm$		11\\
40018-01-02-11	&	5.51$\pm$			0.26&	0.035$\pm$		0.009&	0.520$\pm$		0.007&	1763$\pm$		11\\
40018-01-02-23&	6.56$\pm$		0.23&	0.021	$\pm$		0.005&	0.550$\pm$		0.007&	1720$\pm$		11\\
40018-01-02-10&	7.85$\pm$			0.20&	0.020$\pm$		0.005&	0.590$\pm$		0.007&	1659$\pm$		11\\
\hline\hline
{\bf Cyg X-2/1999}  \\
\hline
40017-02-01-00&	10.57$\pm$	0.06&	0.011$\pm$	0.003&	0.62$\pm$	0.010&	983$\pm$	9\\
40017-02-02-00&	4.03$\pm$	0.22&	0.050$\pm$	0.012&	0.59$\pm$	0.009&	1114$\pm$	9\\
40017-02-01-01&	3.84$\pm$	0.36&	0.024$\pm$	0.006&	0.55$\pm$	0.009&	1074$\pm$	9\\
40017-02-02-01&	4.54$\pm$	0.16&	0.025$\pm$	0.006&	0.60$\pm$	0.009&	1128$\pm$	9\\
40017-02-02-02&	5.61$\pm$		0.20&	0.028$\pm$	0.007&	0.64$\pm$	0.010&	1161$\pm$	9\\
40017-02-03-00&	3.47$\pm$	0.14&	0.020$\pm$	0.005&	0.50$\pm$	0.008&	1236$\pm$	10\\
40017-02-03-01&	3.76$\pm$	0.20&	0.042$\pm$	0.011&	0.50$\pm$	0.008&	1144$\pm$	9\\
40017-02-04-01&	3.39$\pm$	0.28&	0.077$\pm$	0.019&	0.46$\pm$	0.007&	1146$\pm$	9\\
40017-02-04-00&	3.85$\pm$	0.12&	0.042$\pm$	0.011&	0.51$\pm$		0.007&	1310$\pm$	10\\
40017-02-05-00&	8.67$\pm$	0.08&	0.018$\pm$	0.004&	0.40$\pm$	0.006&	1350$\pm$	10\\
\hline\hline
\end{tabular}
\end{table*}

\begin{table*}
	\centering
	\caption{RMS and MTS for BH LMXBs. Full version of the table is available online.}
	\label{tab:example_table}
	\begin{tabular}{rrrrrrrrr} 
	\hline
Obs. Id. & RMS &   $\tau$   & Hardness & PCU2 \\
	\hline
&  $\%$ &  sec & & cnts/sec \\
\hline\hline
{\bf XTE J1859+226/1999}  \\
\hline
40124-01-04-00	& 	29.7	$\pm$ 	0.6	& 	0.033	$\pm$ 	0.006& 0.65 $\pm$	0.01	&726	$\pm$8\\
40124-01-05-00	& 	25.4	$\pm$ 	0.6	& 	0.025	$\pm$ 	0.005&0.45 $\pm$	0.01	& 991$\pm$9	\\
40124-01-06-00	& 	23.5	$\pm$ 	0.3	& 	0.017 	$\pm$ 	0.003&0.42 $\pm$	0.01	&1091$\pm$9	\\
40124-01-07-00	& 	23.6	$\pm$ 	0.4	& 	0.021		$\pm$ 	0.004&0.42 $\pm$	0.01	&1078$\pm$9	\\
40124-01-08-00	& 	21.5	$\pm$ 	0.2	& 	0.016	$\pm$ 	0.003&0.38 $\pm$	0.01	&1193$\pm$ 9\\
40124-01-09-00	& 	20.0	$\pm$ 	0.4	& 	0.019	$\pm$ 	0.004&0.35 $\pm$	0.01	&1268$\pm$10	\\
40124-01-10-00	& 	16.8	$\pm$ 	0.3	& 	0.020	$\pm$ 	0.004&0.33 $\pm$	0.01	&1421$\pm$10	\\
40124-01-11-00		& 	16.5	$\pm$ 	0.3	& 	0.021		$\pm$ 	0.004&0.32 $\pm$	0.01	&1465$\pm$10	\\
40124-01-12-00	& 	3.3	$\pm$ 	0.3	& 	0.009	$\pm$ 	0.002&0.24 $\pm$	0.01	&2666$\pm$14\\
40124-01-13-00	& 	4.3	$\pm$ 	0.3	& 	0.021		$\pm$ 	0.004&0.25 $\pm$	0.01	&2547$\pm$13	\\
\hline\hline
{\bf MAXIJ1659-152/2010}  \\
\hline
95358-01-02-00& 	23.41$\pm$	0.64& 	0.040$\pm$ 	0.008& 	0.65$\pm$ 	0.02& 	445$\pm$ 	6\\
95358-01-02-01& 	23.33$\pm$ 	1.63& 	0.033$\pm$ 	0.006& 	0.59$\pm$ 	0.01& 	413$\pm$ 	6\\
95358-01-02-02& 	23.33$\pm$ 	2.65& 	0.050$\pm$ 	0.010& 	0.52$\pm$ 	0.01& 	496$\pm$ 	6\\
95358-01-03-00& 	21.93$\pm$ 	1.05& 	0.046$\pm$ 	0.009& 	0.52$\pm$ 	0.01& 	490$\pm$ 	6\\
95108-01-02-00& 	23.07$\pm$ 	2.25& 	0.038$\pm$ 	0.007& 	0.53$\pm$ 	0.01& 	498$\pm$ 	6\\
95358-01-03-01& 	21.88$\pm$ 	1.67& 	0.043$\pm$ 	0.008& 	0.50$\pm$ 	0.01& 	431$\pm$ 	6\\
95108-01-03-00& 	21.77	$\pm$ 	1.52& 	0.033$\pm$ 	0.006& 	0.50$\pm$ 	0.01& 	469$\pm$ 	6\\
95108-01-04-00& 	21.52	$\pm$ 	1.21& 	0.030$\pm$ 	0.006& 	0.49$\pm$ 	0.01& 	488$\pm$ 	6\\
95108-01-05-00& 	22.19$\pm$ 	3.77& 	0.037$\pm$ 	0.007& 	0.47$\pm$ 	0.01& 	470$\pm$ 	6\\
95358-01-03-02& 	21.47	$\pm$ 	1.83& 	0.034$\pm$ 	0.006& 	0.44$\pm$ 	0.01& 	499$\pm$ 	6\\
\hline\hline
{\bf H1743-322/2003}  \\
\hline
80138-01-01-00&	27.73$\pm$	1.75&	0.095$\pm$	0.024&	1.00$\pm$	0.047&	108$\pm$		3\\
80138-01-06-00&	18.51	$\pm$	0.45&	0.030$\pm$	0.008&	0.46$\pm$	0.007&	1203$\pm$	9\\
80138-01-07-00&	13.32$\pm$	0.51&	0.024$\pm$	0.006&	0.35$\pm$	0.006&	1150$\pm$	9\\
80146-01-01-00&	12.10$\pm$	0.33&	0.019$\pm$	0.005&	0.34$\pm$	0.004&	1905$\pm$	12\\
80146-01-02-00&	14.65$\pm$	0.34&	0.019$\pm$	0.005&	0.37$\pm$	0.005&	1514$\pm$	10\\
80146-01-03-00&	16.23$\pm$	0.84&	0.031$\pm$	0.008&	0.40$\pm$	0.008&	687$\pm$	7\\
80146-01-04-00&	5.83$\pm$	0.70&	0.035$\pm$	0.009&	0.38$\pm$	0.005&	2201$\pm$	12\\
80146-01-05-00&	5.92$\pm$	0.38&	0.028$\pm$	0.007&	0.39$\pm$	0.004&	2355$\pm$	13\\
80146-01-06-00&	10.91$\pm$	0.18&	0.043$\pm$	0.011&	0.35$\pm$	0.004&	2032	$\pm$	12\\
80146-01-07-00&	9.11$\pm$		0.51&	0.077$\pm$	0.019&	0.36$\pm$	0.005&	1593$\pm$	11\\
\hline\hline
\end{tabular}
\end{table*}



\begin{thebibliography}{}

\bibitem[Abry et al.(2003)]{Abry03} Abry, P., et al.\ 2003, \emph{Self-similarity and long-range dependence through the wavelet lens}, Theory and Applications of Long-Range Dependence, Boston: Birkhauser, 527-556
\bibitem[Addison(2002)]{Addison02} Addison P. S., \ 2002, \emph{The Illustrated Wavelet Transform Handbook}, IOP Publishing Ltd.
\bibitem[Agrawal et al.(2018)]{2018MNRAS.477.5437A} Agrawal, V. K.; Nandi, Anuj; Girish, V.; Ramadevi, M. C., 2018, MNRAS, 477, 5437
\bibitem[Agrawal et al.(2020)]{2020Ap&SS.365...41A} Agrawal, V. K.; Nandi, Anuj; Ramadevi, M. C., 2020, Ap\&SS, 365, 41
\bibitem[Belloni \& Hasinger(1990)]{1990A&A...230..103B} Belloni, T., \& Hasinger, G.\ 1990, \aap, 230, 103
\bibitem[Belloni et al.(2005)]{2005A&A...440..207B}Belloni, T.; Homan, J.; Casella, P. et al., 2005, A\&A, 440, 207 
\bibitem[Belloni et al.(2006)]{2006MNRAS.367.1113B} Belloni, T., Parolin, I., Del Santo, M., et al.\ 2006, \mnras, 367, 1113
\bibitem[Belloni et al.(2011)]{2011BASI...39..409B} Belloni, T.~M., Motta, S.~E., \& Mu{\~n}oz-Darias, T.\ 2011, Bulletin of the Astronomical Society of India, 39, 409
\bibitem[Coifman(1995)]{Coifman95translation-invariantde-noising} Coifman R. R. and Donoho D. L.\ 1995, Translation-invariant de-noising, Springer-Verlag, 125--150
\bibitem[Daubechies \& Heil(1992)]{1992ComPh...6..697D}Daubechies, I.; Heil, C., 1992, ComPh, 6, 697
\bibitem[Done \& Gierlinski(2003)]{2003MNRAS.342..361N} Done, C.; Gierliński, M. \ 2003, MNRAS, 342,1041–1055 
\bibitem[Done et. al.(2007)]{2007A&ARv..15....1D}Done, C.; Gierliński, M.; Kubota, A., 2007,The Astronomy and Astrophysics Review, Volume 15, Issue 1, pp.1-66
\bibitem[Dunn et al.(2010)]{2010MNRAS.403...61D}Dunn, R. J. H.; Fender, R. P.; Körding, E. G. et al. 2010, MNRAS, 403, 61
\bibitem[Fender et al.(2004)]{2004MNRAS.355.1105M}Fender, R. P.; Belloni, T. M.; Gallo, E., 2004, MNRAS, 355, 1105  
\bibitem[Flandrin(1989)]{Flandrin89} Flandrin P.\ 1989, IEEE, Transactions on Information Theory, 35, 197--199
\bibitem[Flandrin(1992)]{Flandrin92} Flandrin P.\ 1992, IEEE, Transactions on Information Theory, 38, 910--917
\bibitem[Gardenier \& Uttley(2018)]{2018MNRAS.481.3761G} Gardenier, D. W. \& Uttley, P. \ 2018, MNRAS, 481, 3761
\bibitem[Gierlinski \& Done (2002)]{2002MNRAS.331.47}Gierlinski, M. \& Done, C., 2002, MNRAS, 331, L47
\bibitem[Gleissner et al.(2004)]{2004A&A...414.1091G}Gleissner, T.; Wilms, J.; Pottschmidt, K.; Uttley, P.; Nowak, M. A.; Staubert, R. 2004, A\&A, 414, 1091
\bibitem[Hasinger \& van der Klis(1989)]{1989A&A...225...79H} Hasinger, G., \& van der Klis, M.\ 1989, \aap, 225, 79
\bibitem[Homan et al.(2001)]{2001ApJS..132..377H} Homan, J., Wijnands, R., van der Klis, M., et al.\ 2001, \apjs, 132, 377
\bibitem[Homan \& Belloni(2005)]{2005Ap&SS.300..107H}Homan, J.; Belloni, T., 2005, Astrophysics and Space Science, Volume 300, Issue 1-3, pp. 107-117
\bibitem[Homan et al.(2010)]{2010ApJ...719..201H} Homan, J.; van der Klis, M.; Fridriksson, J. K. et al. 2010, ApJ,  719, 201   
\bibitem[Jernigan et al.(2000)]{2000ApJ...530..875J} Jernigan, J.~G., Klein, R.~I., \& Arons, J.\ 2000, \apj, 530, 875
\bibitem[Lachowicz \& Czerny (2005)]{2005MNRAS.361..645L} Lachowicz, P. \& Czerny, B., 2005, MNRAS, 361, 645
\bibitem[Lachowicz \& Done (2010)]{2010A&A...515..65L}Lachowicz, P. \& Done, C., 2010, A\&A, 515, A65 
\bibitem[Leahy et al.(1983)]{1983ApJ...266..160L} Leahy, D.~A., Darbro, W., Elsner, R.~F., et al.\ 1983, \apj, 266, 160
\bibitem[Lewin et al.(1988)]{1988SSRv...46..273L} Lewin, W.~H.~G., van Paradijs, J., \& van der Klis, M.\ 1988, \ssr, 46, 273
\bibitem[Lin, Remillard \& Homan(2009)]{2009ApJ...696.1257L} Lin, D.; Remillard, R. A. \& Homan, J. 2009, ApJ, 696, 1257 
\bibitem[Lin, Remillard \& Homan(2010)]{2010ApJ...719.1350L}Lin, D.; Remillard, R. A. \& Homan, J.  2010, ApJ, 719, 1350
\bibitem[Maccarone \& Coppi(2003)]{2003MNRAS.338..189M}Maccarone, T. J. \& Coppi, P. S. 2003, MNRAS, 338, 189
\bibitem[MacLachlan et al.(2013a)]{2013MNRAS.432..857M}MacLachlan, G. A., Shenoy, A., Sonbas, E. et al. 2013a, MNRAS, 432, 857
\bibitem[MacLachlan et al.(2013b)]{2013MNRAS.436.2907M}MacLachlan, G. A., Shenoy, A., Sonbas, E. et al. 2013b, MNRAS, 436, 2907
\bibitem[Mallat(1989)]{Mallat89} Mallat S. G.\ 1989, IEEE, Transactions on Pattern Analysis and Machine Intelligence, 11, 674--693
\bibitem[Mancuso et al.(2021)]{2021MNRAS.502.1856M} Mancuso, G. C.; Altamirano, D.; Méndez, M.; Lyu, M.; Combi, J. A., 2021, MNRAS, 502, 1856 
\bibitem[Miyamoto et al.(1992)]{1992ApJ...391L..21M}Miyamoto, S.; Kitamoto, S.; Iga, S.; Negoro, H., \& Terada, K. 1992, ApJ, 391, L21
\bibitem[Miyamoto et al.(1993)]{1993ApJ...403L..39M} Miyamoto, S., Iga, S., Kitamoto, S., et al.\ 1993, \apjl, 403, L39
\bibitem[Mohamed et al.(2021)]{2021arXiv210104201M} Mohamed, K., Sonbas, E., Dhuga, K.S., Göğüş, E., Tuncer, A., Abd Allah, N.N., Ibrahim, A.,  2021, MNRAS.tmpL, 6
\bibitem[Mu{\~n}oz-Darias et al.(2014)]{2014MNRAS.443.3270M} Mu{\~n}oz-Darias, T., Fender, R.~P., Motta, S.~E., et al.\ 2014, \mnras, 443, 3270
\bibitem[Muno et al.(2002)]{2002ApJ...568L..35M}Muno, Michael P.; Remillard, Ronald A.; Chakrabarty, D., 2002, ApJ, 568, L35
\bibitem[Nandi et al.(2012)]{2012A&A...542A..56N} Nandi, A.; Debnath, D.; Mandal, S. and Chakrabarti, S. K., 2012, A\&A, 542, 56
\bibitem[Nowak et al.(1999)]{1999ApJ...510..874N}Nowak, M. A.; Vaughan, B. A.; Wilms, J.; Dove, J. B.; Begelman, M. C., 1999, ApJ, 510, 874 
\bibitem[Nowak (2000)]{2000MNRAS.318..361N} Nowak, M.~A.\ 2000, \mnras, 318, 361
\bibitem[Percival(2002)]{Percival00} Percival, D. B. and Walden, A. T. \ 2002, Wavelet Methods for Time Series Analysis, Cambridge University Press
\bibitem[Phillipson et al.(2018)]{2018MNRAS.477.5220P} Phillipson, R. A.; Boyd, P. T.; Smale, A. P., 2018, MNRAS, 477, 5220 
\bibitem[Pottschmidt et al.(2003)]{2003A&A...407.1039P}Pottschmidt, K.; Wilms, J.; Nowak, M. A.; Pooley, G. G.; Gleissner, T. et al. 2003, A\&A, 407, 1039 
\bibitem[Remillard \& McClintock(2006)]{2006ARA&A..44}Remillard, R. A., McClintock, J. E., 2006, Annual Review of Astronomy $\&$ Astrophysics, 44, 49
\bibitem[Scargle et al. (1993)]{1993ApJ...411L..91S} Scargle, J.D., et al., 1993, ApJ, 411, L91
\bibitem[Scargle et al. (2013)]{2013ApJ...764..167S} Scargle, J.D., et al., 2013, ApJ, 764, 167
\bibitem[Sonbas et al.(2020)]{2020MNRAS.853.150S}Sonbas, E.; Mohamed, K.; Dhuga, K. S.; Tuncer, A;,Gogus, E., 2020, \mnras, 499, 2513
\bibitem[Sreehari \& Nandi(2021)]{2021MNRAS.502.1334S} Sreehari, H. \& Nandi, A., 2021, MNRAS, 502, 1334
\bibitem[Steiman-Cameron et al. (1997)]{1997ApJ...487..396S}Steiman-Cameron, T.Y., et al., 1997, ApJ, 487, 396
\bibitem[Tao et al.(2018)]{2018MNRAS.480.4443T} Tao, Lian; Chen, YuPeng; Güngör, Can; Huang, Yue et al. 2018, \mnras, 480, 4443
\bibitem[van der Klis(1989)]{1989ARA&A..27..517V} van der Klis, M.\ 1989, \araa, 27, 517
\bibitem[van der Klis(2006)]{2006csxs.book...39V}van der Klis M., 2006, in Lewin W. H. G., van der Klis M., Cambridge Astrophysics Series, No. 39, Compact Stellar X-ray Sources. Cambridge Univ. Press, Cambridge, p. 39
\bibitem[Zhang et al.(1995)]{1995ApJ...449..930Z}Zhang, W.; Jahoda, K.; Swank, J. H.; Morgan, E. H.; Giles, A. B., 1995, ApJ, 449, 930
\bibitem[Zhang et al.(1996)]{1996ApJ...469L..29Z} Zhang, W., Morgan, E.~H., Jahoda, K., et al.\ 1996, \apjl, 469, L29
\end{thebibliography}

\label{lastpage}
\end{document}